\documentclass{elsart}

\usepackage{epsfig}
\usepackage{amssymb}
\usepackage{amsmath}

\begin{document}

\begin{frontmatter}

\title{Approach to a non-equilibrium steady state}

\author[uw]{ Jaros\l aw Piasecki}
\author[uch,ucm]{Rodrigo Soto}
\ead{rsoto@dfi.uchile.cl}
\address[uw]{Institute of Theoretical Physics, University of
Warsaw, Hoza 69, 00 681 Warsaw, Poland}
\address[uch]{Departamento de F\'{\i}sica, Facultad de Ciencias 
F\'\i sicas y Matem\'aticas, Universidad de Chile, Casilla 487-3,
	Santiago, Chile}
\address[ucm]{Departamento de F\'{\i}sica Aplicada I (Termolog\'\i a),
Facultad de Ciencias F\'{\i}sicas, Universidad Complutense, 
28040 Madrid, Spain} 

\begin{abstract}
We consider a non-interacting one-dimensional gas accelerated by a constant
and uniform external field.  The energy absorbed from the field is
transferred via elastic collisions to a bath of scattering obstacles.
At gas-obstacle encounters the particles of the gas acquire a fixed kinetic
energy.  The approach to the resulting stationary state is studied within
the Boltzmann kinetic theory. It is shown that the long time behavior is
governed by the hydrodynamic mode of diffusion superposed on a convective
flow. The diffusion coefficient is analytically calculated for any value of
the field showing a minimum at intermediate field intensities. It is
checked that the properly generalized Green-Kubo formula applies in 
the non-equilibrium stationary state. 
\end{abstract}

\begin{keyword}
 \PACS  05.70.Ln 
   \sep 47.70.Nd 
   \sep 51.10.+y 
   \sep 51.20.+d 
  
\end{keyword}
\end{frontmatter}

\section{Introduction} \label{sec1}
   
  The problem of constructing adequate statistical ensembles for
non-equilibrium stationary states (hereafter denoted by NESS) has a long
history (see e.g. \cite{ZUB,PR}).  Related to it there are very interesting
findings concerning the nature of fluctuations in the presence of spatial
gradients inducing stationary flows \cite{TK}. The computation of
exact expressions for the probability distribution of simple NESS
allows to study different static properties as, for example, the long range
density correlations \cite{DLS}. 
The recent development of
the theory of fluidized granular matter has also attracted a lot of
attention to the structure of NESS (see for example \cite{JK,GG,SPM}). 
Finally, there is at present an important research
carried on in order to properly define the entropy production accompanying
stationary dissipative currents (see e.g. \cite{PG}).  
    
However, whereas the modes of approach of fluids to thermal equilibrium are
well understood, there is still no satisfactory theory of the dynamics of
reaching NESS. In both cases the final state is stationary, but in NESS, in
contradistinction to equilibrium, there persist non-vanishing fluxes
sustained by the coupling to environment.

The kinetic theory of fluids revealed the fundamental role of the
separation between different time scales in the approach to equilibrium 
(see e.g. the presentation of Bogoliubov's ideas in \cite{AP}).  In
particular, in the evolution of the density of particles in the
one-particle phase space all the degrees of freedom but those represented
by the densities of globally conserved quantities
relax at a short time
scale producing a state close to local equilibrium. The subsequent slow
evolution is essentially governed by changes of the hydrodynamic fields (an
original construction of the hydrodynamic modes within the kinetic theory
has been developed by P. R\'esibois \cite{PRes}). One can wonder whether
this kind of mechanism based on the time-scale separation still persists
when the final state involves some dissipative stationary process related
to the current flowing through the system. This is precisely the question
we want to address in this paper.

Our object here is to exploit an analytically soluble one-dimensional model
discussed in \cite{Pias86} in order to perform a detailed study of the
evolution towards a NESS. Although the system is very simple indeed, we
find that the possibility of a rigorous study of its dynamics is precious
from the point of view of the search for the laws governing the approach to
NESS. The coupling to the environment is represented in the model by the
action of an external field. One of the original results of the paper is an
explicit formula for the field dependent coefficient of diffusion governing
the evolution of the only hydrodynamic mode in the approach to the
stationary state.  

In Sec. \ref{sec2} we define the system and its dynamics. The following
analysis of the approach to the stationary state is presented in Sec.
\ref{sec3}. Finally, Sec. \ref{sec4} contains the concluding comments.

\section{One-dimensional system and its dynamics}\label{sec2}

   We consider non-interacting point particles of mass $m$ moving in one
dimension with acceleration $a$ under the influence of an external
constant and uniform field. The particles are surrounded by a bath of point
obstacles of mass $m$ that move with velocities $\pm U$, both directions
being equally probable.  At elastic collisions between the particles and
the obstacles a simple interchange of velocities is taking place.  The
scattering obstacles are uniformly distributed with a number density $n$.

  The statistical state of the particles is described by the probability
density $f(r,v,t)$ for finding a particle at time $t$ at point $r$ with
velocity $v$. We assume that the evolution of the density $f(r,v;t)$  is
governed by the linear Boltzmann equation \cite{Pias86} 
\begin{equation}
\left( \frac{\partial}{\partial t}+v \frac{\partial}{\partial r}+
a\frac{\partial}{\partial v} \right) f(r,v,t) = n \int
dc |v-c| \left[ f(r,c,t) \phi(v)-f(r,v,t) \phi(c) \right] ,
\label{ke}
\end{equation}
putting
\begin{equation}
 \phi(v)=\frac{1}{2}\left[ \delta (v+U)+\delta (v-U) \right] 
\label{phi}
\end{equation}
for the velocity distribution of the obstacles. When writing equation
(\ref{ke}) it has been supposed that the particles always encountered
obstacles with velocities distributed according to  (\ref{phi}). The
possibility of recollisions with obstacles perturbed by previous collisions
with the particles is not taken into account by the Boltzmann collision
term in (\ref{ke}).

In dimensionless variables $w=v/U$, $x=nr$, and $\tau=Unt $ equation (\ref{ke}) takes the form
\begin{align}
\left( \frac{\partial}{\partial \tau}+w\frac{\partial}{\partial x}
+\epsilon\frac{\partial}{\partial w} +
\frac{1}{2}(|w+1|+|w-1|)\right) F(x,w,\tau) =& \nonumber\\
\quad \frac{1}{2}[
\delta(w-1)\mu_{-}(x,\tau)+\delta(w+1)\mu_{+}(x,\tau) ] ,  &\label{dke}
\end{align}
where  $F(x,w,\tau)=f(x/n,Uw,\tau/nU)U/n$ is the dimensionless probability
density, and
\begin{equation}
 \mu_{\pm}(x,\tau)=\int\,dw |w\pm 1| F(x,w,\tau) . \label{mu}
\end{equation}
In equation (\ref{dke}) there appears the dimensionless intensity of the
field $\epsilon=a/nU^2$.

Provided that appropriate boundary conditions are supplied at
$\pm\infty$, equation (\ref{dke}) has a non-equilibrium stationary solution
representing
a homogeneous NESS whose velocity distribution $F_0(w;\epsilon)$, that can 
be chosen to be normalized as  $\int dw F_0(w;\epsilon)=1$, has been
derived and analyzed in \cite{Pias86}. The NESS is characterized by a
constant current $V_{\rm NESS}(\epsilon)=\int dw w F_0(w;\epsilon)$ that
shows a linear response, $V_{\rm NESS}(\epsilon)\sim\epsilon$, in the
weak field limit $|\epsilon|\ll1$ and a non-linear response, $V_{\rm
NESS}(\epsilon)\sim{\rm sgn}(\epsilon)\sqrt{|\epsilon|}$, for strong fields
$|\epsilon|\gg1$.

Whereas the paper \cite{Pias86} focused on the properties of the stationary
distribution, our aim here is to analyze the approach to the asymptotic
homogeneous NESS starting from an arbitrary inhomogeneous initial
condition. This requires solving the time-dependent Boltzmann equation.
The way towards the construction of the solution has been already found in
\cite{Pias86}. Using the Fourier-Laplace transformation
\begin{equation}
\tilde{F}(k,w,z) = \int_0^{\infty} d\tau \int dx e^{-z \tau- i k x }
F(x,w,\tau)
\end{equation}
one finds from (\ref{dke}) the integral equation
\begin{align}
\tilde F(k,w,z) &= \tilde H(k,w,z) + \frac{1}{2|\epsilon
|}\exp\left[-\frac{\chi (w)}{\epsilon}-ik\frac{w^2-1}{2\epsilon}\right]
\nonumber\\
&\quad \times \left\{ \exp\left[ \frac{1-z(w-1)}{\epsilon}\right] \theta
[\epsilon (w-1)]\tilde\mu_-(k,z)\right. \nonumber\\
&\left. \quad+ \exp\left[ -\frac{1+z(w+1)}{\epsilon}\right] \theta
[\epsilon (w+1)]\tilde\mu_+(k,z) \right\},
\label{solimplicit}
\end{align}
where
\begin{equation}
\chi(w) = \left[ (w+1)|w+1| + (w-1)|w-1| \right] /4,
\end{equation}
that obey the relation $\chi(w)=-\chi(-w)$,
\begin{equation}
\tilde \mu_\pm(k,z) = \int dw |w\pm1|\tilde F(k,w,z), \label{Defmus}
\end{equation}
and 
\begin{equation}
\tilde H(k,w,z)=\int_0^\infty e^{-z \tau-ikx} H(x,w,\tau)
\label{DefH}
\end{equation}
relates to the initial condition $F(x,w,0)$ through
\begin{equation}
H(x,w,t) = \exp[(\chi(w-\epsilon \tau)-\chi(w))/\epsilon]
F(x-w\tau+\epsilon\tau^2/2,w-\epsilon
\tau,0)   \label{H}
\end{equation}
Eq. (\ref{solimplicit}) is implicit expressing $\tilde F(k,w,z)$  in terms
of $\tilde \mu_\pm(k,z)$ that are linear functionals of $\tilde F(k,w,z)$.
However, the solution can be made explicit in a straightforward way.
Indeed, multiplying Eq. (\ref{solimplicit}) by $|w\pm1|$ and integrating
with respect to $w$ one obtains a closed system of linear equations for
$\tilde \mu_\pm$ in the form
\begin{equation}
{\rm {\bf M}}(k,z;\epsilon ) \cdot \left[ \begin{array}{l} \tilde
\mu_-(k,z) \\ \tilde \mu_+(k,z)\end{array} \right]
=\left[\begin{array}{l} h_-(k,z) \\ h_+(k,z)\end{array} \right] .
\label{Eqmus}
\end{equation}
Here
\begin{equation}
h_\pm(k,z) = \int dw |w\pm1|\tilde H(k,w,z) \label{Defh}
\end{equation}
and ${\rm {\bf M}}(k,z;\epsilon )$ is a two by two matrix analytic in $z$
given by
\begin{equation}
 {\rm {\bf M}}(k,z;\epsilon)= \begin{pmatrix} 
M_{11}(k,z;\epsilon), & M_{12}(k,z;\epsilon)\\
M_{21}(k,z;\epsilon), & M_{22}(k,z;\epsilon)
\end{pmatrix} ,
\end{equation}
with
\begin{align*} 
M_{11}&= \frac{1}{2|\epsilon |} \int dw |w-1|\exp\left[-\frac{\chi
(w)}{\epsilon}-ik\frac{w^2-1}{2\epsilon}\right]
 \exp\left[ \frac{1-z(w-1)}{\epsilon}\right] \theta [\epsilon (w-1)]\\
M_{12}&=  \frac{1}{2|\epsilon |}\int dw |w-1| \exp\left[-\frac{\chi
(w)}{\epsilon}-ik\frac{w^2-1}{2\epsilon}\right]
\exp\left[ -\frac{1+z(w+1)}{\epsilon}\right] \theta [\epsilon (w+1)]\\
M_{21}&= \frac{1}{2|\epsilon |} \int dw |w+1|\exp\left[-\frac{\chi
(w)}{\epsilon}-ik\frac{w^2-1}{2\epsilon}\right]
 \exp\left[ \frac{1-z(w-1)}{\epsilon}\right] \theta [\epsilon (w-1)]\\
M_{22}&=  \frac{1}{2|\epsilon |}\int dw |w+1| \exp\left[-\frac{\chi
(w)}{\epsilon}-ik\frac{w^2-1}{2\epsilon}\right]
\exp\left[ -\frac{1+z(w+1)}{\epsilon}\right] \theta [\epsilon (w+1)].
\end{align*}

When $\epsilon >0$, the matrix ${\rm {\bf M}}(k,z;\epsilon )$ takes the
form
\begin{equation}
{\rm {\bf M}}(k,z;\epsilon )=
\begin{pmatrix} 
 1-(A-B)e^{(2z+1)/2\epsilon}  ,& C-D+(B-A)e^{-(2z+3)/2\epsilon}  \\
-(A+B)e^{(2z+1)/2\epsilon}, & 1-(A+B)e^{-(2z+3)/2\epsilon}-C-D 
\end{pmatrix} ,
\end{equation}
where
\begin{align}
A(k,z;\epsilon) &= \frac{1}{2\epsilon} \int_1^\infty dw\, w
\exp\left\{-\frac{1}{2\epsilon}\left[ w^2(1+ik) + 2wz -ik\right] \right\}\\
B(k,z;\epsilon) &= \frac{1}{2\epsilon} \int_1^\infty dw\,
\exp\left\{-\frac{1}{2\epsilon}\left[ w^2(1+ik) + 2wz -ik\right] \right\}\\
C(k,z;\epsilon) &= \frac{1}{2\epsilon} \int_{-1}^1 dw\, w
\exp\left\{-\frac{1}{2\epsilon}\left[ ikw^2 + 2(w+1)(z+1)-ik\right]
\right\} \\
D(k,z;\epsilon) &= \frac{1}{2\epsilon} \int_{-1}^1 dw\,
\exp\left\{-\frac{1}{2\epsilon}\left[ ikw^2 + 2(w+1)(z+1)-ik\right]
\right\} .
\end{align}
The corresponding formulae for $\epsilon <0$ follow from the important
symmetry relations
\begin{align}
M_{11}(k,z;\epsilon)&= M_{22}(-k,z;-\epsilon)\\
M_{12}(k,z;\epsilon)&= M_{21}(-k,z;-\epsilon). \label{symm}
\end{align}

\section{Modes of Approach to the Non-Equilibrium Steady State}\label{sec3}

The linear equation (\ref{Eqmus}) contains all information about the modes
by which the system approaches the NESS. Whether these modes  behave for
long wavelengths as hydrodynamic modes is the question we want to study
now. To this end we use the explicit solution of the Boltzmann equation.
The structure of (\ref{solimplicit}) and (\ref{Eqmus}) implies that $\tilde
F(k,w,z)$ is an analytic function in the complex plane except at points $z$
where the inverse matrix ${\rm {\bf M}}^{-1}$ does not exist. Thus the
zeros of the determinant ${\rm Det}({\rm {\bf M}})$ define the
singularities in the $z$-plane of the Fourier-Laplace transform $\tilde F
(k,w,z)$. In particular, isolated zeros $z_i(k)$ corresponding to simple
poles will produce after applying the inverse Laplace transformation an
exponential time dependence of the form
\begin{equation}
\hat F(k,w,\tau) = \sum_i \alpha_i\, e^{z_i(k)\tau} ,
\end{equation}
with coefficients $\alpha_i$ for each mode depending on the initial
condition ($\hat F$ denotes the spatial Fourier transform of $F$).

We performed a systematic numerical survey of the zeros of the
determinant ${\rm Det}({\rm {\bf M}})$ for different values of $k$ and
$\epsilon$. It has been found that regardless of the value of $\epsilon$
there is a single isolated zero which behaves like $z_0=-Dk^2-ikV+{O}(k^3)$
for $k\ll1$ .
All the other zeroes, both for finite $k$ and in the limit $k\to 0$, have
negative real parts and are located outside a band of a certain finite
width around the imaginary axis. That is, these other zeros do not have an
accumulation point with vanishing real part.
Therefore, for
long wavelengths, there is exactly one slow diffusive mode (coefficient
$D$) combined with convective transport with velocity $V$. All the others
are fast kinetic modes. The presence of an unique slow mode is related to
the fact that the mass is the only conserved quantity. No zeroes with
positive real part were found, reflecting the fact that the NESS is stable.

\begin{figure}[htb]
\begin{center}
\epsfig{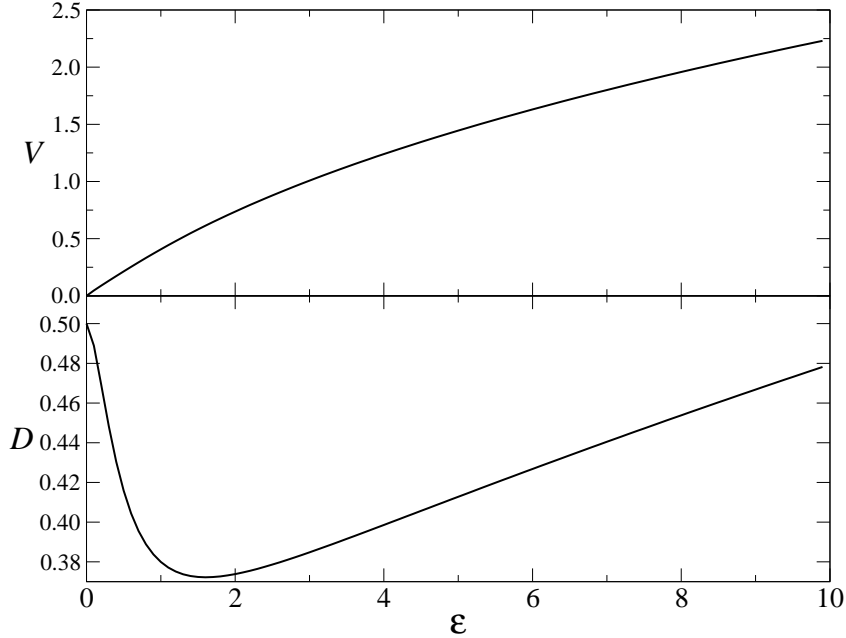}
\end {center}
\caption{Top: Average velocity $V_{\rm NESS}(\epsilon)$ in the NESS  as a
function of the imposed
acceleration $\epsilon$. Bottom: Field dependent diffusion
coefficient $D(\epsilon)$.}
\label{fig.DV}
\end{figure}

    The slow mode can be obtained analytically by inserting the asymptotic
formula $z=-D(\epsilon )k^2-ikV(\epsilon) $ into the equation  ${\rm
Det}({\rm {\bf M}})=0$ and
then solving for $D$ and $V$ keeping only dominant terms in $k$. 
The relations (\ref{symm}) imply that the zeros of the determinant of the
matrix
${\bf M}(k,z; \epsilon )$  coincide with  the zeros of the determinant of
the matrix
${\bf M}(-k,z; -\epsilon )$. It follows the symmetry relations
\begin{align}
D(\epsilon) &= D(-\epsilon)  = {\mathcal D}(|\epsilon|), \\
V(\epsilon) &= -V(-\epsilon) = {\rm sgn}(\epsilon){\mathcal
V}(|\epsilon|). \label{symmrel}
\end{align}
One finds then a unique solution
\begin{equation}
{\mathcal V}(\epsilon) = \frac{\epsilon [1+e^{-2/\epsilon}] + [\epsilon -1
-(1+\epsilon)e^{-2/\epsilon}]I(\epsilon )}{\epsilon
[1-e^{-2/\epsilon}] + [-\epsilon +3 +(1+\epsilon
)e^{-2/\epsilon}]I(\epsilon )} \label{Vepsilon}
\end{equation}
and
\begin{align}
{\mathcal D}(\epsilon) &=
\bigg[
e^{2/\epsilon} \big( -2 \epsilon^3 + \epsilon^4 + 2 \epsilon^5 + 
     8 \epsilon^2 I(\epsilon) + 2 \epsilon^3 I(\epsilon) -
     10 \epsilon^4 I(\epsilon) - 8 \epsilon^5 I(\epsilon) - 
     10 \epsilon {I(\epsilon)}^2 \nonumber\\
&  -11 \epsilon^2 {I(\epsilon)}^2 - 
     2 \epsilon^3 {I(\epsilon)}^2 + 7 \epsilon^4 {I(\epsilon)}^2 +
     6 \epsilon^5 {I(\epsilon)}^2 + 
     4 {I(\epsilon)}^3 + 8 \epsilon {I(\epsilon)}^3 +
     12 \epsilon^2 {I(\epsilon)}^3 \nonumber\\
&  +14 \epsilon^3 {I(\epsilon)}^3 + 6 \epsilon^4 {I(\epsilon)}^3 \big)
+ e^{4/\epsilon}\big( -4\epsilon^3 + 2 \epsilon^4 - 4 \epsilon^5
    +20 \epsilon^2 I(\epsilon) - 8 \epsilon^3 I(\epsilon) \nonumber\\
&  -12 \epsilon^4 I(\epsilon) + 16 \epsilon^5 I(\epsilon) 
    -20\epsilon {I(\epsilon)}^2 - 2 \epsilon^2 {I(\epsilon)}^2 + 
     20 \epsilon^3 {I(\epsilon)}^2 + 22 \epsilon^4 {I(\epsilon)}^2 -
     12 \epsilon^5 {I(\epsilon)}^2 \nonumber \\
&   + 4 {I(\epsilon)}^3 + 16 \epsilon^2 {I(\epsilon)}^3 +
     8 \epsilon^3 {I(\epsilon)}^3 -12 \epsilon^4 {I(\epsilon)}^3
     \big)
+  e^{6/\epsilon} \big( -2 \epsilon^3 - 3 \epsilon^4 + 2 \epsilon^5
\nonumber \\
&   + 12 \epsilon^2 I(\epsilon) - 26 \epsilon^3 I(\epsilon) +
22 \epsilon^4 I(\epsilon) - 
     8 \epsilon^5 I(\epsilon) - 10 \epsilon {I(\epsilon)}^2 -
3 \epsilon^2 {I(\epsilon)}^2 + 
     30 \epsilon^3 {I(\epsilon)}^2 \nonumber\\
&   - 29 \epsilon^4 {I(\epsilon)}^2 + 6 \epsilon^5 {I(\epsilon)}^2 - 
     8 \epsilon {I(\epsilon)}^3 + 20 \epsilon^2 {I(\epsilon)}^3 -
22 \epsilon^3 {I(\epsilon)}^3 + 
     6 \epsilon^4 {I(\epsilon)}^3 \big)  \bigg] \nonumber \\
&/\bigg[\epsilon\big(\epsilon - I(\epsilon) - \epsilon I(\epsilon) +
e^{2/\epsilon} 
   \left( -\epsilon - 3 I(\epsilon) + \epsilon I(\epsilon)
\right)\big)^3\bigg] , \label{Depsilon}
\end{align}
with
\begin{equation}
I(\epsilon) = e^{1/2|\epsilon|} \int_1^\infty dw\, e^{-w^2/2|\epsilon|} .
\end{equation}
The drift velocity $V(\epsilon)$ (\ref{Vepsilon}) coincides, as expected,
with the stationary average velocity $V_{\rm NESS}$  found in
\cite{Pias86}. In
Fig. \ref{fig.DV} we plotted its dependence on $\epsilon$, showing the
transition from a linear to a nonlinear response.  

The explicit expression for the diffusion coefficient (\ref{Depsilon}) is
quite involved. A plot of it is presented in Fig. \ref{fig.DV}. It should
be noticed that $D(\epsilon)$ has an interesting structure as a function of
the intensity of the external field showing a minimum for
$\epsilon\approx1.60$. The asymptotic formulae for small and large
intensities are

\begin{alignat}{2}
D(\epsilon) &\approx \frac{1}{2} -\frac{15}{8}\epsilon^2 &; \quad
|\epsilon|&\ll
1 ,\\
D(\epsilon) &\approx \frac{4-\pi}{\sqrt{2\pi^3}} \epsilon^{1/2} +
\frac{9\pi-20}{(2\pi)^{3/2}} \epsilon^{-1/2}   &; \quad
\epsilon&\gg 1 .\label{squareroot}
\end{alignat}

It should be remarked that in the limit of vanishing external field the   
value $D(0)$ coincides with that following at equilibrium
($\epsilon=0$) from the Green-Kubo formula. For large values of $\epsilon$,
the velocity of the obstacles becomes negligible compared to that acquired
by accelerated motion  and their action resembles that of stopping centers.
Therefore by purely dimensional analysis one can predict the exponent $1/2$
in the dependence of $D(\epsilon)$ in (\ref{squareroot}).
 
\begin{figure}[htb]
\epsfclipon
\begin{center}\epsfig{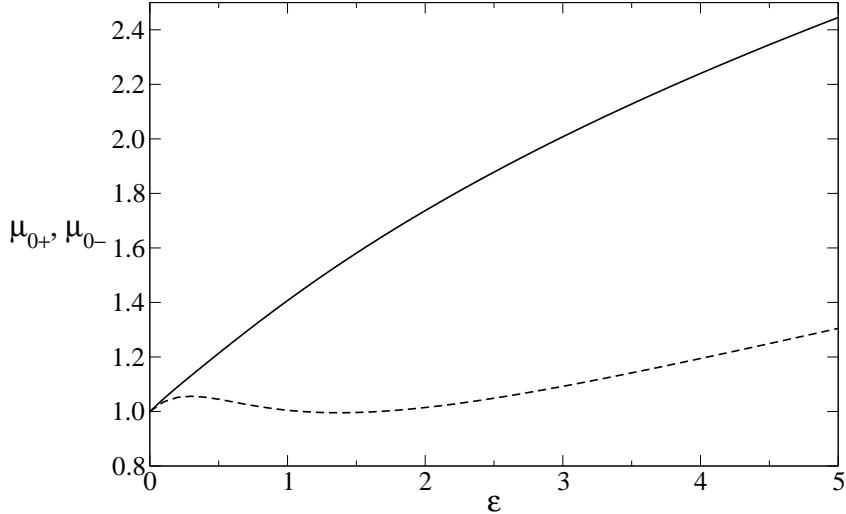}
\end{center}
\caption{Partial collision frequencies in the NESS, solid line
$\mu_{0+}$, dashed line $\mu_{0-}$.}
\label{fig.Mu}
\end{figure}

  The appearance of a minimum in $D(\epsilon)$ shown in Fig. \ref{fig.DV}
is an interesting point which calls for interpretation. The minimum can be
understood by analyzing the partial collision frequencies with obstacles
moving with velocities $+1$ or $-1$. In the NESS, these collision
frequencies are given by 
\begin{equation}
\mu_{0\pm}(\epsilon) = \int dw |w\pm 1|F_{0}(w;\epsilon) ,
\end{equation}
where the average is computed with the use of the NESS distribution
$F_0(w;\epsilon)$ given in Eq.(40) of Ref. \cite{Pias86}. As shown in Fig.
\ref{fig.Mu}, the collision frequency $\mu_{0-}$ with obstacles moving to
the right (in the direction of the accelerating field) has a minimum at
$\epsilon=1.37$. In order to understand this fact consider a particle that
has just collided with an obstacle with velocity $-1$ getting
instantaneously its velocity. If the field is weak, it will continue moving
to the left encountering obstacles with velocity $+1$ and will be sent
again towards those with velocity $-1$ the collision frequency $\mu_{0-}$
remaining high. Also if the field is large enough, the particle will turn
rapidly to the right and gain a large velocity leading again to a large
value of $\mu_{0-}$. There are however intermediate values of acceleration
where the particle remains a long time turning to the right under the
action of the field  before the next collision occurs without getting large
velocities and thus reducing the collision rate. This fact explains the
origin of the minimum in $\mu_{0-}(\epsilon)$. Finally, the presence of the
minimum in $\mu_{0-}$ can be related to the observed minimum of
$D(\epsilon)$ because the lowering of the collision rate for a range of
values of $\epsilon$ produces an evolution that is closer to a
deterministic one, and thus less diffusive.

Knowing the analytic solution of the initial value problem for the kinetic
equation (\ref{dke}) we could also verify  if the diffusion coefficient
(\ref{Depsilon}) follows from the Green-Kubo formula when the NESS
distribution is used in the evaluation of the autocorrelation function. We
thus considered the Green-Kubo expression
\begin{align}
D_{\rm GK} &=\lim_{z\to0} \int_0^\infty d\tau e^{-z\tau}  
\langle (w-V)(\tau) (w-V)(0)\rangle_{\rm NESS}\\
&= \lim_{z\to0} \int_0^\infty d\tau e^{-z\tau} 
\int dw (w-V) F_{\rm GK}(w,\tau) ,
\end{align}
where $F_{\rm GK}(w,\tau)$ is the solution of the equation (\ref{dke}) with
initial condition $F_{\rm GK}(w,0)=(w-V)F_0(w)$ and $V$ is the NESS drift
velocity. The computation process is as follows. Given the initial
condition $F_{\rm GK}(w,0)$, the equation (\ref{Eqmus}) is solved for
$\tilde
\mu_\pm$ (note that $z$ must be nonzero because otherwise the inverse
matrix ${\rm {\bf M}}(k=0,z)^{-1}$ does not exist). Then,
$\tilde{F}_{\rm GK}(w,z)$ is obtained and the coefficient $D_{\rm GK}$ can
be found
from the relation
\begin{equation}
D_{\rm GK} =\lim_{z\to0} \int dw (w-V) \tilde{F}_{\rm GK}(w,z)
\label{Dif.GK} .
\end{equation}
We performed the calculation obtaining a finite value which coincides with
(\ref{Depsilon}). We thus conclude that when the system is in the NESS, both
in the linear and in the nonlinear response regimes, the diffusion
coefficient is given by the Green-Kubo formula provided the NESS
distribution is used in the averaging process and the  fluctuations of the
velocity around the non-zero  mean value $V_{\rm NESS}$ are considered.
This
conclusion is also confirmed by an analogous result obtained for a simpler
one-dimensional case where only nonlinear transport is present
\cite{Pias83}.

Finally, we have checked that beyond the linear response regime 
the system does not obey the
fluctuation-dissipation theorem, relating the diffusion coefficient $D$ and
the mobility $\mu$. The mobility is defined as $\mu=d V_{\rm
NESS}/d\epsilon$. Evaluating the non-equilibrium stationary temperature 
$T=\langle (w-V_{\rm NESS})^2\rangle$ using the stationary
distribution $F_0$, it is directly checked that $D(\epsilon)\neq
T(\epsilon)\mu(\epsilon)$. The equality is only satisfied at the
equilibrium case $\epsilon=0$. A similar phenomenon has been predicted in
granular gases \cite{DG01}.

\section{Concluding comments}\label{sec4}

We have considered a one-dimensional system of particles absorbing energy
from a constant and uniform external field, and dissipating it via
collisions with a bath of scattering obstacles with a fixed velocity
distribution (\ref{phi}). In this situation the appropriate linear Boltzmann
equation (\ref{dke}) predicts the approach to a non-equilibrium steady state
(NESS) characterized by a mass flux that can present  both linear and
nonlinear response depending on the strength of the field. It has been
found that the evolution of arbitrary initial conditions toward the NESS
involves short time scale kinetic modes producing the stationary velocity
distribution followed by  the slow long wavelength hydrodynamic diffusion 
superposed on a convective current flowing with the average velocity of the
NESS.

Analytic calculations show an interesting structure in the dependence of
the diffusion coefficient  on the intensity of the external field, with a
minimum at intermediate values of the field. The minimum is related to the
presence of another minimum in the collision frequency between the
particles  and the obstacles, thus reducing the number of randomizing
scattering processes. Furthermore, we have also shown that the value of the
diffusion coefficient could be obtained from a Green-Kubo formula by
considering the time displaced peculiar velocity correlation function
(subtracting the average NESS velocity) averaged over the NESS
distribution.

     It thus appears that the modes of approach to the NESS when only the
mass remains as a conserved quantity  involve the corresponding single
hydrodynamic mode which is the classical process of spatial diffusion. The
diffusion coefficient does depend on the external field intensity and can
be obtained
at any value of the field from an appropriately generalized Green-Kubo
formula.
We conclude that the approach to NESS  shows
qualitative analogy to that of reaching equilibrium. Hopefully this
conclusion remains valid for a large class of systems in which the
stationary state results from the balance between the energy flow absorbed
from an external field and collisional coupling to a kind of thermostat.

The authors thank P. Cordero for helpful discussions. This work has been
partly financed by {\em Fondecyt} research grants 1030993 and 7040123 and
{\em Fondap} grant 11980002.

\end{document}